\documentclass[12pt,twoside,fleqn]{article}

\usepackage{epsfig}
\usepackage{psfig}

\def\lsim{\raise0.3ex\hbox{$<$\kern-0.75em\raise-1.1ex\hbox{$\sim$}}}
\def\gsim{\raise0.3ex\hbox{$>$\kern-0.75em\raise-1.1ex\hbox{$\sim$}}}
\makeatletter
\@addtoreset{equation}{section}
\makeatother

\newcommand{\beqn} {\begin{equation}}
\newcommand{\eqn} {\end{equation}}

\newcommand{\slsh}[1] {#1\kern-.43em/}

\newcommand{\real}{{\sf I}\kern-.12em{\sf R}}
\newcommand{\comp}{{\sf I}\kern-.48em{\sf C}}
\newcommand{\nin} {\in\kern-.6em/}

\def\MEF{m_{\rm eff}}\def\mef{\ifmmode\MEF\else$\MEF$\fi}

\def\NT{N_\tau}
\def\nt{\ifmmode\NT\else$\NT$\fi}
\def\NS{N_\sigma}
\def\ns{\ifmmode\NS\else$\NS$\fi}

\def\PL{{ Phys.\ Lett.\ }}
\def\NP{{ Nucl.\ Phys.\ }}

\def\MO{{\langle |L| \rangle }}

\def\EN{{\langle P \rangle }}
\def\ENQ{{\langle P^2 \rangle }}

\begin{document}
\thispagestyle{empty}
%
 \mbox{} \hfill BI-TP 96/48~~\\
 \mbox{} \hfill October 1996\\
\begin{center}
\vspace*{1.0cm}
{{\large \bf The Pseudo Specific Heat in SU(2) Gauge Theory :\\
         Finite Size Dependence and Finite Temperature Effects}
 } \\
\vspace*{1.0cm}
{\large J. Engels and T. Scheideler } \\
\vspace*{1.0cm}
{\normalsize
$\mbox{}$ {Fakult\"at f\"ur Physik, Universit\"at Bielefeld,
D-33615 Bielefeld, Germany}}\\
\vspace*{2cm}
{\large \bf Abstract}
\end{center}
\setlength{\baselineskip}{1.3\baselineskip}

We investigate the pseudo specific heat of SU(2) gauge theory near the 
crossover point on $4^4$ to $16^4$ lattices. Several different methods 
are used to determine the specific heat. 
The curious finite size dependence of the peak maximum 
is explained from the interplay of the 
crossover phenomenon with the deconfinement
transition occurring due to the finite extension of the lattice. 
We find, that for lattices of size $8^4$ and larger the crossover peak 
is independent of lattice size at $\beta_{co}=2.23(2)$ and has a peak 
height of $C_{V,co}=1.685(10)$. We conclude therefore that the crossover 
peak is not the result of an ordinary phase transition. Further, the 
contributions to $C_V$ from different plaquette correlations 
are calculated. We find, that at the peak and far outside the peak 
the ratio of contributions from 
orthogonal and parallel plaquette correlations is different.
To estimate the finite temperature influence on symmetric lattices 
even far off the deconfinement transition point we calculate 
the modulus of the lattice average of the 
Polyakov loop on these lattices and compare it to predictions from a
random walk model.

\newpage

\setcounter{page}{1}
\vspace*{1cm}

\section{Introduction}


The pseudo specific heat $C_V$ of $SU(2)$ gauge theory was already 
investigated in the beginning of Monte Carlo lattice studies.
It is known to have a peak near $\beta=4/g^2 \approx 2.2$, in the
crossover region between strong and weak coupling behaviour.
A first finite size analysis by Brower et al. \cite{Brow81} on $4^4-10^4$
lattices revealed a strange dependence of the peak on the volume
$V=(N_{\sigma}a)^4$ of the lattice. Here, $a$ is the lattice spacing
and $N_{\sigma}$ the number of points in each direction. The location 
of the peak shifts with increasing volume from smaller $\beta-$values
to larger and then to smaller ones again; the peak maximum decreases
with increasing volume. Such a behaviour is unknown for any ordinary
phase transition. The nature and origin of the peak remained therefore
unclear, though a connection to the nearby endpoint of the first order
critical line in the ($\beta, \beta-$adjoint)-plane was proposed by
Bhanot and Creutz \cite{Bhan81}: "The peak in the specific heat of the
$SU(2)$ model is a shadow of this nearby singularity".
Recently, new interest in this extended $SU(2)$ model came up again 
\cite{Gava95,Step96} and a line of second order deconfinement transition
points was found, connecting the normal deconfinement transition point 
at $\beta_A =0$ to the endpoint of the first order critical line
inside the ($\beta, \beta_A$)-plane. The origin of the change from 
second to first order behaviour and the possibility of a bulk transition
have still to be examined.

For both the relation to the extended $SU(2)$ model and the unexplained 
finite size behaviour in the crossover region a new study of
the pseudo specific heat is useful. Moreover,
we may now determine $C_V$ with much higher statistics and also on larger 
lattices than in the early calculations, and we can apply
modern analysis techniques.
In addition, since symmetric lattices are often used to simulate
zero temperature physics, it is important to estimate remaining finite
temperature effects which may show up in the pseudo specific heat .


\section{Methods to Calculate $C_V$}


We use the standard Wilson action for $SU(2)$
\begin{equation}
S = \beta \cdot \sum_{x,\mu \nu} P_{\mu \nu}(x)~,
\end{equation}
where 
\begin{equation}
P_{\mu \nu}(x) = 1 - {1 \over 2}{\rm Tr}U_{\mu \nu}(x)~,
\end{equation}
is the plaquette or energy and $U_{\mu \nu}(x)$ is the plaquette
link operator. The sum extends over all independent forward plaquettes.
There are $N_P = 6N_{\sigma}^4$ such plaquettes. We denote the
lattice average of the plaquettes by $P$
\begin{equation}
P = {1 \over N_P} \sum_{x,\mu \nu} P_{\mu \nu}(x)~.
\end{equation}
The speudo specific heat is then defined by
\begin{equation}
C_V = { d\EN \over d(1/\beta) } = - \beta^2 { d\EN \over d\beta }~.
\end{equation}
There are three methods to determine $C_V$ :
\par \noindent
i) one measures the plaquette expectation values $\EN$ as a function
of $\beta$ and calculates the numerical derivative at $\beta_M=\beta+
\Delta\beta/2$ from
\beqn
C_V(\beta_M) = 
-{\beta_M^2 \over \Delta\beta} (\EN(\beta+\Delta\beta)
-\EN(\beta))~;
\eqn
ii) one measures the variance of the plaquettes, which is proportional
to $C_V$
\beqn
C_V = \beta^2 N_P (\ENQ-\EN^2)~;
\eqn
or,
\begin{figure}[htb]
\begin{center}
   \epsfig{bbllx=85,bblly=230,bburx=495,bbury=560,
       file=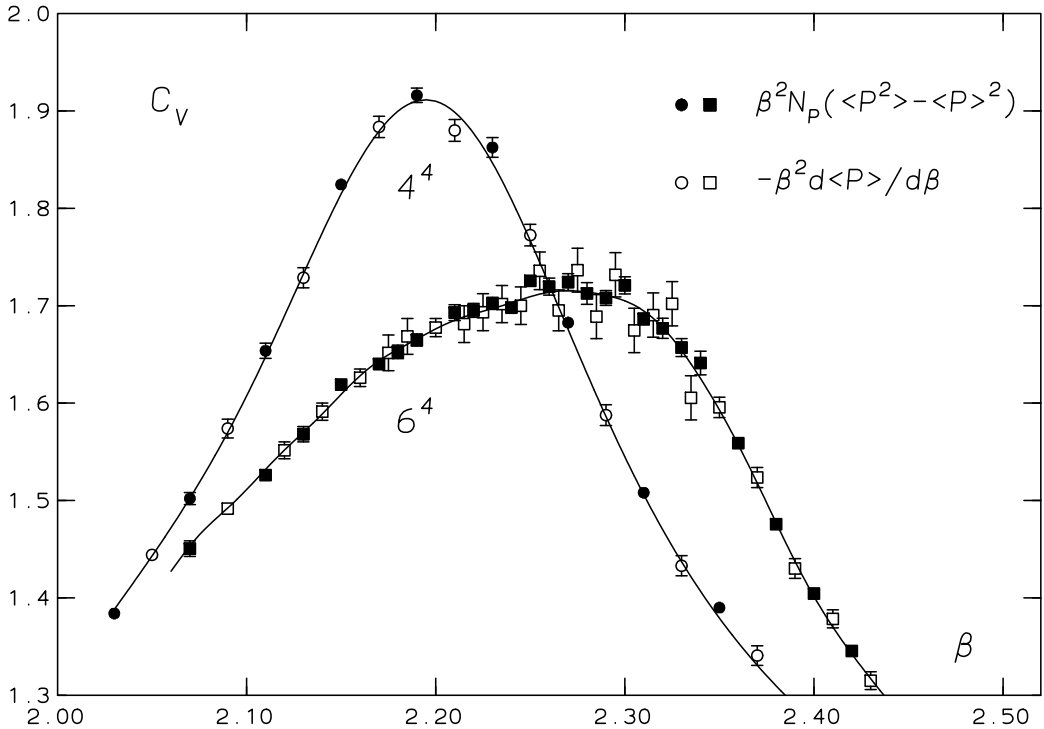, width=100mm,height=80mm, angle=-90}
\end{center}
\caption{The pseudo specific heat $C_V$ calculated from the variance
(filled symbols), the numerical derivative (open symbols) and the 
DSM interpolation (lines), calculated on a $4^4$ (circles) 
and a $6^4$ lattice (squares). }

\label{fig:cvn6}
\end{figure}

\par\noindent
iii) one calculates the sum of plaquette-plaquette correlations
\beqn
C_V = \beta^2 \sum_{x^{\prime},\mu^{\prime} \nu^{\prime}} (
\langle P_{\mu \nu}(x) P_{\mu^{\prime} \nu^{\prime}}(x^{\prime}) \rangle
-\EN^2)~.
\label{meth3}
\eqn
The most straightforward way is, of course, to calculate the 
variance of $\EN$. The density of states method (DSM) \cite{DSM} may then be
used to interpolate between the points. With increasing lattice size
the DSM requires however more and more simulation points to obtain a
densely populated action histogram. This comes about, because the 
variance of $P$ is essentially proportional to $N_P^{-1}$ ( implying
a nearly $N_P-$independent specific heat $C_V$, except for the smallest 
lattices ). From
\beqn
S = \beta N_P P~,
\eqn
it follows then, that the width of an action histogram for a fixed 
$\beta-$value is $\sim N_P^{1/2}$. 
Compared to that the size of the complete multiple points histogram for the 
action is of order $N_P$. Thus more simulation points are needed for
larger lattices to interpolate in the same $\beta-$range. Consequently
we have only applied the DSM to the $4^4$ and $6^4$ lattices.
 
In Fig.~\ref{fig:cvn6} we show the results from methods 
i) and ii) and the DSM interpolation for these lattices. At each $\beta-$value
we took on the average 90-120 thousand measurements. 
Between the measurements five updates, 
consisting of one heatbath and two overrelaxation steps were performed,
so that the autocorrelation time was of the order of one.
As can be seen from Fig.~\ref{fig:cvn6} there is complete consistence 
between the different methods. It is perhaps appropriate 
at this point to note that if, as it is often done, 
the plaquettes are measured during and not
after each update the resulting plaquette variance is about $20\%$ smaller 
than expected, because of local correlations among the plaquettes, though
the plaquette average is correct.

\begin{figure}[htb]
\begin{center}
   \epsfig{bbllx=85,bblly=230,bburx=495,bbury=560,
       file=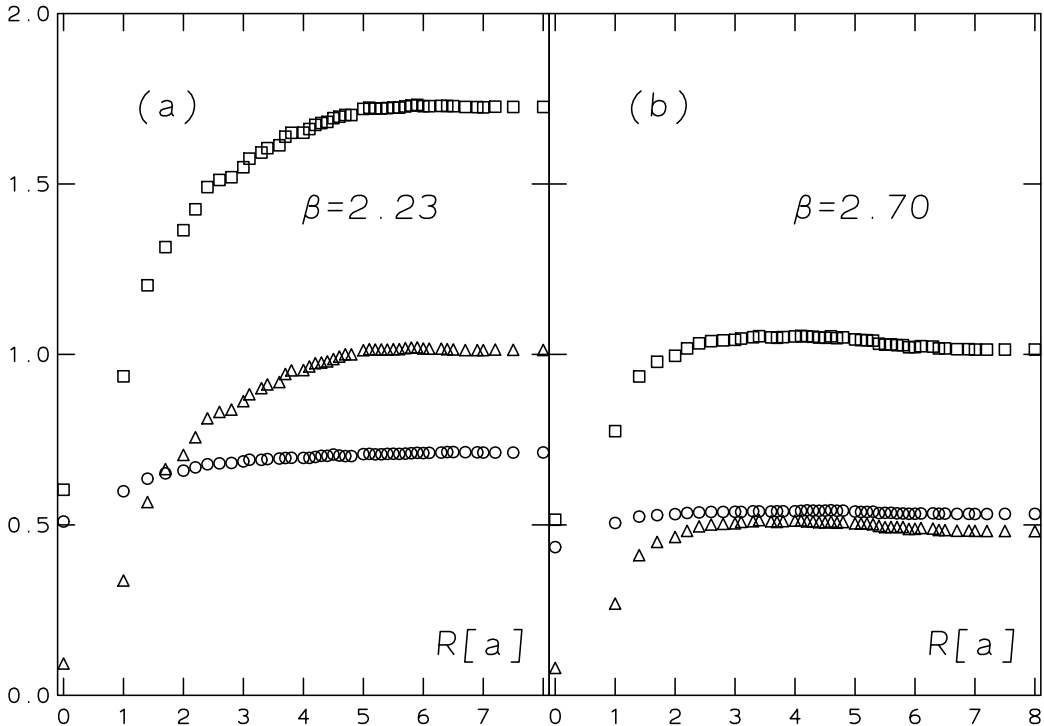, width=100mm,height=80mm, angle=-90}
\end{center}
\caption{Contributions to $C_V$ on an $8^4$ lattice from parallel (circles) 
and orthogonal (triangles) plaquette correlations up to distance $R$ as a 
function of $R$. The sum is plotted as squares. The results are shown  
for $\beta=2.23$ (a) and $\beta=2.70$ (b).}

\label{fig:corr}
\end{figure}
 
We have also investigated the plaquette correlations. We find in general 
a rapid fall with $R=x^{\prime}-x$, the correlation length is of order 1-2
lattice spacings.
The plaquettes $P_{\mu \nu}(x)$ and $P_{\mu^{\prime} \nu^{\prime}}(x^{\prime})$ 
in Eq. (\ref{meth3})
may be in parallel or orthogonal planes. At the peak ($\beta \approx 2.23$)
we find that the total contribution of the orthogonal correlations is
about $30\%$ higher than that of the parallel correlations, whereas far away 
from the peak, at $\beta=2.70$, the contributions are essentially equal.
This is demonstrated in Fig. \ref{fig:corr}, where we compare the 
contributions to $C_V$ in Eq. (\ref{meth3}). Shown are the corresponding sums
up to distance $R$ as a function of $R$. Note, that the maximal nontrivial 
(with respect to the periodic boundary conditions) diagonal distance $R$ 
on an $\ns^4$ lattice is $\ns a$. We see that at the crossover the
orthogonal correlations reach their total contribution only at $R \approx 4a$,
whereas at $\beta=2.70$ a distance of $R \approx 2a$ is sufficient. In contrast
to that the parallel correlations have a much shorter range and distances
of $R > 1.5a$ play no role, both at the crossover and at higher $\beta-$values. 
There is also no difference among those parallel correlations where the two
plaquettes are in the same plane or in different parallel planes.

\section{Finite Size Dependence of $C_V$ and Finite Temperature Effects}

\begin{figure}[htb]
\begin{center}
   \epsfig{bbllx=110,bblly=200,bburx=470,bbury=590,
       file=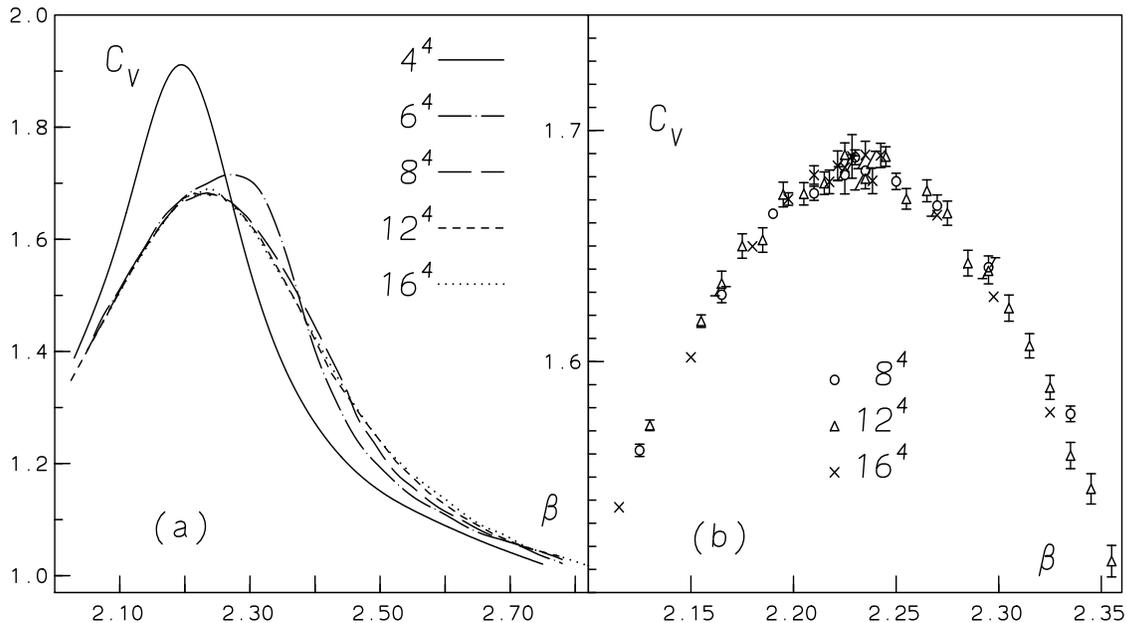, width=85mm,height=85mm,angle=-90}
\end{center}
\caption{The pseudo specific heat $C_V$ vs. $\beta$ on $\NS^4$ lattices.
Part (a) shows the results for $\ns=4$ and 6 from the DSM interpolation,
for $\ns=8,12$ and 16 the measured points were connected by lines to guide
the eye. Part (b) shows the crossover region for $\ns=8,12$ and 16.}   
\label{fig:sizeall}
\end{figure}
\begin{figure}[htb]
\begin{center}
   \epsfig{bbllx=80,bblly=195,bburx=525,bbury=610,
       file=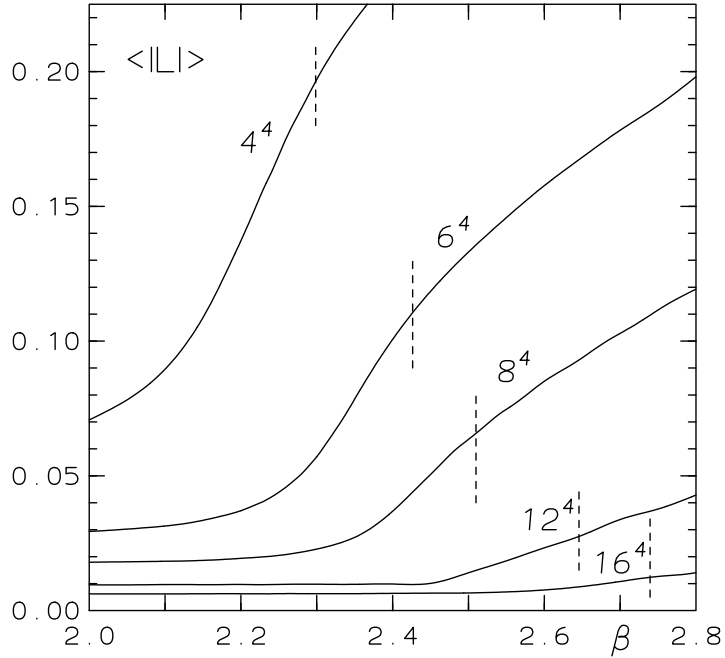, height=90mm}
\end{center}
\caption{The expectation value of the modulus of $L$ for $\NS=4,6,8,12$
and 16 vs. $\beta$. The broken vertical lines show the locations of the 
corresponding finite temperature phase transitions. }

\label{fig:Poly}
\end{figure}

In Fig. \ref{fig:sizeall}(a) we compare the results for $C_V$ from lattices
with $\ns=4,6,8,12$ and 16. The general behaviour already found in
\cite{Brow81} is fully confirmed. On the other hand, we see that there
is no further finite size dependence in the peak region, if $\NS \geq 8$.
This is shown in Fig. \ref{fig:sizeall}(b) in more detail for the crossover 
region.The results suggest, that the finite size dependence 
of the smaller lattices is related to a 
different phenomenon. Indeed, the critical point for the $\nt=4$ finite 
temperature deconfinement transition is at $\beta_c = 2.30$, very close to the
crossover peak positon. Since we are using periodic boundary conditions
for all directions, the approach to the critical point corresponding to
$\ns$ will influence the plaquette expectation values. To check this, we
have calculated the Polyakov loop 
\beqn
L(\vec{x}) =  {1\over 2}{\rm Tr}
\prod_{t=1}^{\NS} U_{ \vec{x},t }~~,
\eqn
and its lattice average $L$ 
\beqn
L = {1\over N_\sigma^3} \sum_{\vec{x}} L(\vec{x})~~.
\eqn
\newpage
\noindent
As can be seen from Fig. \ref{fig:Poly}, the expectation value of the
modulus of $L$ is not zero on symmetric lattices, not even in the strong 
coupling limit, i.e. we have finite temperature effects also on symmetric
lattices. Well below $\beta_c(\NS)$ the quantity $\langle |L| \rangle$
is a constant. With increasing $\beta$ it starts to increase already before
the transition point. It is obvious, that due to the nearby transition
points the crossover peaks of the $4^4$ and $6^4$ lattices are more
distorted than those of the larger lattices, where only the right shoulders
of the peaks are slightly influenced. 

We can actually calculate the $\NS-$dependence
of $\MO$ at $\beta=0$ from a simple random walk model. For $\beta=0$
the Polyakov loops $L(\vec{x})$ at the $\ns^3$ positions $\vec{x}$
are a set of equal random variables. The expectation value of 
the modulus of their sum is then assumed to behave as 

\beqn
\langle |\sum_{\vec{x}} L(\vec{x})| \rangle \sim (\ns^3)^{1/2}~~,
\eqn
\noindent
for large $\ns$, so that
\beqn
\MO_{\beta=0} = c \cdot \ns^{-3/2}~~,
\label{strong}
\eqn
\noindent
where $c$ is a constant. 
Indeed, a simulation at $\beta=0$ confirms this relation in detail.  

\begin{figure}[htb]
\begin{center}
   \epsfig{bbllx=80,bblly=195,bburx=525,bbury=610,
       file=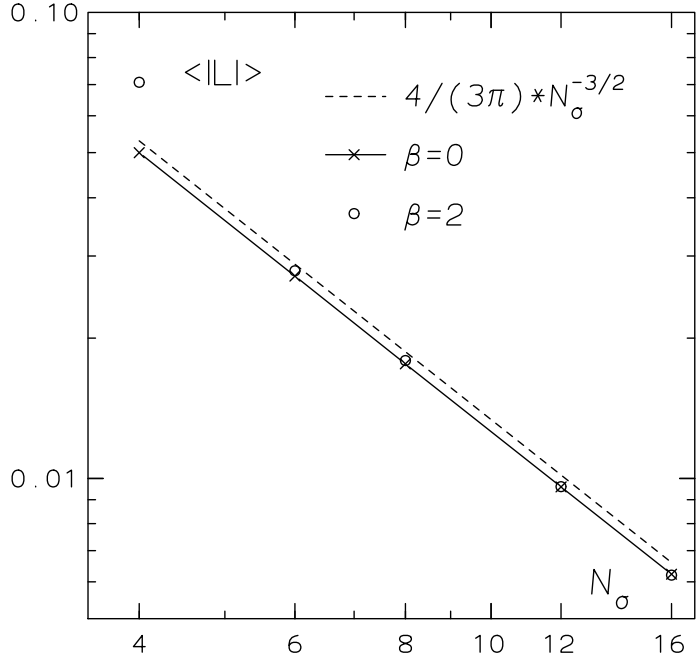, height=90mm}
\end{center}
\caption{The logarithm of $\langle |L| \rangle$ vs. ln$\ns$ on $4^4-16^4$
lattices at $\beta=0$ (solid line), at $\beta=2$ ( circles ) and from 
an estimate from the $\langle |L| \rangle$ value at $\ns=1$ (dashed line). }

\label{fig:logl}
\end{figure}

\noindent
The logarithms of the $\MO -$values, which we calculated at $\beta=0$  
are shown in Fig. \ref{fig:logl} together with the corresponding results 
for $\beta=2$~.
At this $\beta-$value all lattices apart from the $4^4$ lattice have 
already reached their respective strong coupling limits. Deviations
from the limit indicate then the onset of finite temperature effects.
One may thus test, in which $\beta-$region results from a specific 
symmetric lattice can be used to simulate zero temperature physics.
For $SU(2)$ a fit to the $\beta=0$ data leads to $c=0.400(1)$ . 
The relation Eq. (\ref{strong}) holds as well in $SU(3)$ gauge theory and we 
find there the value $c=0.296(2)$. A first estimate of 
the proportionality constant $c$ is obtained from the expectation value
of the modulus of a single $L(\vec{x})$, i. e. the result for $\ns=1$.
For $SU(2)$ it is
\beqn
\langle |L(\vec{x})| \rangle_{\beta=0} = 4/3\pi~~,
\eqn
which differs from $c$ only by $5\%$. 

 Our final conclusion is, that the crossover peak is not
the result of an ordinary phase transition. For large lattices
the peak is at $\beta_{co}=2.23(2)$, its height is
$C_{V,co}=1.685(10)$. For small lattices $(\ns \leq 6)$ however, 
the interplay of the crossover phenomenon and finite 
temperature effects shift and distort the peak considerably. 
\medskip
\noindent


{\large \bf Acknowledgements}


\noindent
The work has been supported by the Deutsche Forschungsgemeinschaft 
under grant Pe 340/3-3. We thank F. Karsch for stimulating discussions.



\end{document}